\documentclass[%
aps,%
prb,%
reprint,%
amsmath,amssymb,%
floatfix,%
nofootinbib,%
superscriptaddress%
]{revtex4-2}

\usepackage[T1]{fontenc}
\usepackage[latin9]{inputenc}
\usepackage[english]{babel}
\usepackage{mathrsfs}
\usepackage{bbm}
\usepackage{graphicx}
\usepackage{xcolor}
\usepackage{physics}
\usepackage{bm}
\usepackage[export]{adjustbox}

\usepackage[colorlinks,
            linkcolor=blue,    
            citecolor=blue,  
            urlcolor=blue,
 	        bookmarks=true,        
	        bookmarksopen=true,    
	        bookmarksnumbered=true,
]{hyperref}

\newcommand{\av}[1]{\left\langle#1\right\rangle}

\newcommand{\id}{\mathbbm{1}}

\begin{document}

\title{Exact NESS of XXZ circuits boundary driven with arbitrary resets or fields}

\author{Vladislav Popkov}

\affiliation{Department of Physics, Faculty of Mathematics and Physics, University of Ljubljana, Jadranska 19, SI-1000 Ljubljana, Slovenia}
 \affiliation{Department of Physics, University of
  Wuppertal, Gaussstra\ss e 20, 42119 Wuppertal, Germany}

\author{Toma\v z Prosen}
%\email{tomaz.prosen@fmf.uni-lj.si}

\affiliation{Department of Physics, Faculty of Mathematics and Physics, University of Ljubljana, Jadranska 19, SI-1000 Ljubljana, Slovenia}
\affiliation{Institute of Mathematics, Physics and Mechanics, Jadranska 19, SI-1000 Ljubljana, Slovenia}

\begin{abstract}
We propose spatially inhomogeneous matrix product ansatz for
an exact many-body density operator of a boundary driven
XXZ quantum circuit.
The ansatz has formally infinite bond-dimension and is fundamentally different from previous constructions.
The circuit is driven by a pair of reset quantum channels applied on the boundary qubits, which polarize the qubits to arbitrary pure target states. Moreover, one of the reset channels can be replaced by an arbitrary local unitary gate, thus representing a hybrid case with coherent/incoherent driving.  Analyzing the ansatz we obtain a family of relatively robust separable nonequilibrium steady states (NESS), which can be viewed as a circuit extension of spin-helix states, and are particularly suited for experimental investigations.
\end{abstract}

\maketitle

\emph{Introduction.---} 
The theory of open quantum systems~\cite{BreuerPetruccione,NielsenChuang,GardinerZoller} represents the backbone of quantum simulation engineering in the so-called NISQ (Noisy Intermediate Scale Quantum) era.
Having exactly solvable models of strongly correlated (open) quantum systems, in particular of the type of digital quantum circuits, is of unprecedented value for the control, benchmarking and science demonstration purpose of NISQ devices (see e.g.~\cite{GateSet,GQAI_Nature22,GQAI_Science24a,GQAI_Science24b}).

An important paradigm in this context is that of integrable quantum spin chains driven out of equilibrium by dissipative processes at the chain's boundaries~\cite{TopicalReview}.
Most remarkably, it has been shown that the nonequilibrium steady state (NESS) of a trotterized integrable XXZ quantum spin chain, realized as a brickwork quantum circuit, driven by a certain type of Kraus dissipators at the chain ends can be written as an infinite bond-dimension matrix product ansatz (MPA)~\cite{VanicatPRL18}. This protocol has later been simulated by a NISQ device~\cite{GQAI_Science24b}.
The setup not only represents elegant mathematical solution of simple nontrivial correlated many-body model, but it also leads to interesting, and sometimes exotic new physics, such as fractal transport coefficients (Drude weights)~\cite{PRL106,LjubotinaPRL19},
insulating far-from equilibrium spin transport or negative differential conductance~\cite{Benenti,PRL107}.

Recently, a new type of MPA with spatially inhomogeneous matrices have been introduced in order to solve the boundary driven NESS of XXZ/XYZ chain in the Zeno regime of asymptotically large boundary dissipation~\cite{ZadnikPRL2020,ZadnikPRE2020,PopkovPRB2022}. In this Letter we report an unexpected extension of inhomogeneous infinite-bond dimension MPA for expressing a NESS of a brickwork XXZ circuit, driven by reset channels targeting arbitrary boundary states and/or external boundary magnetic field oriented in arbitrary direction. The auxiliary space has a tensor product structure of 
two replicas corresponding to `bra' and `ket' of the density matrix. However, unlike in all previous exactly solvable cases, the boundary reset-channel driving nontrivially couples these two replica spaces. 
In that sense, our ansatz is fundamentally different from previous constructions, where the many-body density operator can be factorized in terms of a product (Cholesky-like decomposition~\cite{TopicalReview}).
Our results thus not only promise immediate applications in the context of NISQ simulators but also hold the potential of extending the correlated two-replica MPA to other driven models. 
Moreover, we demonstrate an existence, for a specific set of boundary parameters, of a family of relatively robust (and separable) NESS, which can be viewed as an extension to circuits of spin-helix states~\cite{PopkovSHS,PopkovSHS2} which turned out to be prominent in cold-atom experiments~\cite{KettNature2022}.

\begin{figure}[tbp]
	\centering
	\includegraphics[width=0.85\columnwidth]{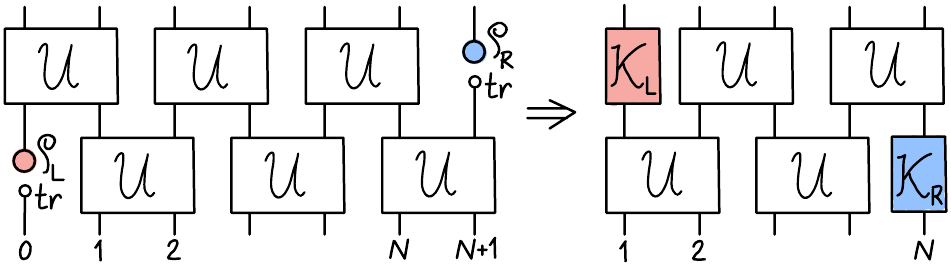}
	\caption{2-step reset driven $XXZ$ circuit in folded notation. Empty circles represent a local traces while red/blue disks represent arbitrary pure boundary states. The right scheme correspond to an equivalent reduced circuit where squares represent boundary Kraus maps (details in text).}
	\label{fig:scheme}
\end{figure}

\emph{Boundary reset driven XXZ circuits.---} We consider a chain of $N+2$ qubits labelled by $0,1\ldots N,N+1$. An XXZ gate, sometimes also referred to as fSim gate~\cite{GateSet,GQAI_Nature22}, can be algebraically parameterized as
\begin{equation}
U=\left(
\begin{array}{cccc}
 1 & 0 & 0 & 0 \\
 0 & a & b & 0 \\
 0 & b& a & 0 \\
 0 & 0 & 0 & 1 \\
\end{array}
\right),\;
a\equiv \frac{q-q^{-1}}{q\lambda\!-\! (q\lambda)^{-1}},\;
b\equiv \frac{\lambda-\lambda^{-1}}{q\lambda\!-\!(q\lambda)^{-1}}.
\label{Ugate}
\end{equation}
$U$ is unitary if: (i) $|q|=1$ and $\lambda\in\mathbb R$
(easy plane regime, EPR), or $(ii)$ $q\in\mathbb R$ and 
$|\lambda|=1$ (easy axis regime, EAR).
Note also $U^{-1}=U|_{q\to q^{-1}} = U|_{\lambda\to \lambda^{-1}}$, $\forall q,\lambda\in\mathbb C$.
The gate acts on a pair of neighboring qubits via so-called folded map $\mathcal U : {\rm End}(\mathbb C^4) \to
{\rm End}(\mathbb C^4)$ as $\mathcal U(\rho) = U \rho U^\dagger$. Assuming $N$ to be odd (while similar construction can be given for $N$ even also), we define a two step dynamical protocol (see Fig.~\ref{fig:scheme}): In step 1, we apply a string of $(N+1)/2$ gates to all neighboring pairs of qubits $1,2\ldots N+1$ while qubit $0$ is traced out and reset to an arbitrary pure state $\rho_{\rm L} = \ket{\psi_{\rm L}}\!\bra{\psi_{\rm L}}$. In step 2, we similarly apply $(N+1)/2$ gates to qubits $0,\ldots N$ and reset the last qubit to arbitrary pure state 
$\rho_{\rm R} = \ket{\psi_{\rm R}}\!\bra{\psi_{\rm R}}$.
This defines a a simple many-body locally interacting (trotterized XXZ) quantum channel, targeting specific local spin states et the ends. Here $\log|\lambda|$ represents the Trotter time step, so $\lambda\to 1$ corresponds to continuous time limit.

Ignoring the trivially separable state of the boundary qubit (left in step 1, and right in step 2), we can write an equivalent effective brickwork dynamical quantum channel acting on interior $N$ qubits, labelled $1,2\ldots N$, as
\begin{eqnarray}
\rho_{t+1} &=& \mathcal M( \rho_t),\,\quad\qquad\qquad \mathcal M = \mathcal M_{\rm o} \mathcal M_{\rm e}, \label{eq:M}\\
\mathcal M_{\rm o} &=& \mathcal K_{\rm L} \otimes \mathcal U^{\otimes (N-1)/2} 
\quad
\mathcal M_{\rm e} = \mathcal U^{\otimes (N-1)/2} \otimes \mathcal K_{\rm R}
, \nonumber
\end{eqnarray}
where ${\mathcal K}_{\rm L/R} : {\rm End}(\mathbb C^2) \to
{\rm End}(\mathbb C^2)$
are the local boundary channels
\begin{equation}
\mathcal K_{\rm L}(\rho) = \tr_1 U (\rho_{\rm L}\otimes \rho)U^\dagger,\;
\mathcal K_{\rm R}(\rho) = \tr_2 U (\rho\otimes \rho_{\rm R})U^\dagger,
\label{eq:reset}
\end{equation}
conveniently expressed in terms of pairs of Kraus matrices
$\mathcal K_{\rm L/R}(\rho) = \sum_{\mu=1}^2 K^{[\mu]\dagger}_{\rm L/R} \rho K^{[\mu]}_{\rm L/R}$:
\begin{eqnarray}
K^{[1]}_{\rm L} &=&
\frac{1}{\sqrt{1+|z|^2}}
\begin{pmatrix}
1 & \frac{\lambda-\lambda^{-1}}{\lambda/q-q/\lambda}\bar{z}
\cr
0 & \frac{q-q^{-1}}{q/\lambda- \lambda/q}
\end{pmatrix},\nonumber \\
K^{[2]}_{\rm L} &=&
\frac{1}{\sqrt{1+|z|^2}}
\begin{pmatrix}
\frac{q-q^{-1}}{q/\lambda-\lambda/q}\bar{z} & 0\cr
\frac{\lambda-\lambda^{-1}}{\lambda/q-q/\lambda} & \bar{z}
\end{pmatrix},\nonumber \\
K^{[1]}_{\rm R} &=&
\frac{1}{\sqrt{1+|w|^2}}
\begin{pmatrix}
1 & \frac{\lambda-\lambda^{-1}}{\lambda/q-q/\lambda}\bar{w}
\cr
0 & \frac{q-q^{-1}}{q/\lambda- \lambda/q}
\end{pmatrix},\nonumber \\
K^{[2]}_{\rm R} &=&
\frac{1}{\sqrt{1+|w|^2}}
\begin{pmatrix}
\frac{q-q^{-1}}{q/\lambda-\lambda/q}\bar{w} & 0\cr
\frac{\lambda-\lambda^{-1}}{\lambda/q-q/\lambda} & \bar{w}
\end{pmatrix}\,.\label{eq:kraus}
\end{eqnarray}
The boundary spinors are stereographically parametrized
\begin{equation}
\ket{\psi_{\rm L}}=\frac{1}{\sqrt{1+|z|^2}}\begin{pmatrix}1 \cr \bar{z}\end{pmatrix}\,,\;
\ket{\psi_{\rm R}}=
\frac{1}{\sqrt{1+|w|^2}}\begin{pmatrix}1 \cr \bar{w}\end{pmatrix}\,,\label{eq:TS}
\end{equation}
in terms of polar and azimuthal angles $z=\tan(\theta_{\rm L}/2)e^{{\rm i}\phi_{\rm L}}$,
$w=\tan(\theta_{\rm R}/2)e^{{\rm i}\phi_{\rm R}}$.
The goal of this Letter is to derive an elegant closed form representation of a unique 
%\textcolor{red}{(the uniqueness is guaranteed by the %contraction mapping theorem)} 
non-equilibrium steady state (NESS) density operator, i.e. fixed point of $\mathcal M$, Eq.~(\ref{eq:M}): $\rho_\infty = \mathcal M (\rho_\infty)$.

\emph{Inhomogeneous Yang-Baxter equation.---}
Let us denote physical space of $N$ qubits as
$\mathfrak h = (\mathbb C^2)^{\otimes N}$ and label local operators over $\mathfrak h$ which act nontrivially only on $n$-th tensor factor (qubit) by subscript $n$, e.g. Pauli operators $\sigma^{\rm x,y,z}_n$. In addition, we consider an infinite-dimensional auxiliary space $\mathfrak a$ spanned by basis vectors $\{\ket{j}_{\mathfrak a};j=0,1\ldots\}$.
We now define a special type of inhomogeneous Lax operators, and their conjugates, $L^\pm_{\mathfrak a,n},L^{\pm*}_{\mathfrak a,n} \in {\rm End}(\mathfrak a\otimes \mathfrak h)$ which we construct as
\begin{eqnarray}
L^+_{\mathfrak a,n} &=& \sum_{j=0}^\infty 
\left[\ket{j}\!\bra{j}_{\mathfrak a} 
A^{[n-2j]}_n D_n
 + 
\ket{j}\!\bra{j\!+\!1}_{\mathfrak a}
A^{[n-2j]}_n E_n\right], \quad \label{eq:Ls} \\
L^-_{\mathfrak a,n} &=& \sum_{j=0}^\infty 
\left[\ket{j}\!\bra{j}_{\mathfrak a} 
E_n A^{[n-2j]}_n 
 + 
\ket{j}\!\bra{j\!+\!1}_{\mathfrak a}
D_n A^{[n-2j]}_n\right], \nonumber \\
L^{+*}_{\mathfrak a,n} &=& \sum_{j=0}^\infty 
\left[\ket{j}\!\bra{j}_{\mathfrak a} 
D_n^\dagger A^{[n-2j]\dagger}_n 
 + 
\ket{j}\!\bra{j\!+\!1}_{\mathfrak a}
E_n^\dagger A^{[n-2j]\dagger}_n\right], \nonumber \\
L^{-*}_{\mathfrak a,n} &=& \sum_{j=0}^\infty 
\left[\ket{j}\!\bra{j}_{\mathfrak a} 
A^{[n-2j]\dagger}_n E^\dagger_n 
 + 
\ket{j}\!\bra{j\!+\!1}_{\mathfrak a}
A^{[n-2j]\dagger}_n D^\dagger_n\right], \nonumber  
\end{eqnarray}
where
\begin{eqnarray}
A^{[n]} &=& 
\begin{pmatrix}
1 & - q^{-n}/z \cr
q^n z & -1
\end{pmatrix}
= \begin{pmatrix} 1\cr q^n z    
\end{pmatrix}(1,-q^{-n}/z)\,, \nonumber\\
D &=& 
\begin{pmatrix}
\lambda & 0 \cr
0 & 1
\end{pmatrix}\,,
\qquad\;
E = 
\begin{pmatrix}
1 & 0 \cr
0 & \lambda
\end{pmatrix}.
\end{eqnarray}
Note that $z\in \mathbb C$ is a free formal parameter here.
Interpreting the gate $U_{n,n+1}$ as a braid-type R-matrix of the unitary six-vertex model,  we postulate the following inhomogeneous (i.e. explicitly $n$-dependent) Yang-Baxter identities (or RLL relations):
\begin{eqnarray}
U_{n,n+1}L^+_{\mathfrak a,n} L^-_{\mathfrak a,n+1} &=& 
L^-_{\mathfrak a,n} L^+_{\mathfrak a,n+1} U_{n,n+1},
\nonumber \\
U_{n,n+1}L^{+*}_{\mathfrak a,n} L^{-*}_{\mathfrak a,n+1} &=& 
L^{-*}_{\mathfrak a,n} L^{+*}_{\mathfrak a,n+1} U_{n,n+1}.
\label{eq:RLL1}
\end{eqnarray}
For proving these identities, and noting that a shift $j\to j+1$ in auxiliary space is equivalent to $z\to q^2 z$, it is enough to establish that they hold for matrix elements, 
$\bra{0}\bullet\ket{0}_{\mathfrak a}$,
$\bra{0}\bullet\ket{1}_{\mathfrak a}$,
$\bra{0}\bullet\ket{2}_{\mathfrak a}$, which is 
straightforward.

\emph{Explicit inhomogeneous matrix product NESS.---}
Let us now define the so-called doubled Lax operators acting over 2-replica auxiliary space,
$\mathbb L^{\pm}_{\mathfrak{ab},n} \in
{\rm End}(\mathfrak a \otimes \mathfrak b\otimes \mathfrak h)$, $\mathfrak b\simeq \mathfrak a$, as
$\mathbb L^\pm_{{\mathfrak ab},n} = 
L^\pm_{\mathfrak a,n}
L^{\pm*}_{\mathfrak b,n}
$ again satisfying RLL identity:
\begin{equation}
U_{n,n+1} 
\mathbb L^+_{\mathfrak{ab},n}
\mathbb L^-_{\mathfrak{ab},n+1}U^\dagger_{n,n+1} =
\mathbb L^-_{\mathfrak{ab},n}
\mathbb L^+_{\mathfrak{ab},n+1}\,.
\label{eq:bulk}
\end{equation}
The NESS fixed point can be conveniently split into even-odd 2-cycle:
\begin{equation}
\rho_\infty' = \mathcal M_{\rm e}(\rho_\infty),\quad
\rho_\infty = \mathcal M_{\rm o}(\rho'_\infty),
\label{eq:fixpoint}
\end{equation}
for which we postulate the matrix product ans\" atze (MPA):
\begin{eqnarray}
\rho_\infty &=& 
 \bra{\rm L}
\mathbb L^+_{\mathfrak{ab},1}\mathbb L^-_{\mathfrak{ab},2} \mathbb L^+_{\mathfrak{ab},3}\cdots
\mathbb L^-_{\mathfrak{ab},N-1}
\mathbb L^+_{\mathfrak{ab},N}
\ket{\rm R}, \nonumber\\
\rho'_\infty &=& 
 \bra{\rm L}
\mathbb L^-_{\mathfrak{ab},1}\mathbb L^+_{\mathfrak{ab},2}\mathbb L^-_{\mathfrak{ab},3}\cdots 
\mathbb L^+_{\mathfrak{ab},N-1}
\mathbb L^-_{\mathfrak{ab},N}
\ket{\rm R},
\label{eq:MPA}
\end{eqnarray}
where
\begin{equation}
\bra{\rm L} = \sum_{j,j'}\ell_{j,j'} \bra{j}_{\mathfrak{a}}\bra{j'}_{\mathfrak{b}}\,,\;\;
\ket{\rm R} = \sum_{j,j'}r_{j,j'} \ket{j}_{\mathfrak{a}}\ket{j'}_{\mathfrak{b}}\,,
\end{equation}
are the boundary vectors as elements of 2-replica auxiliary space. $\ket{\rm L}$ and $\bra{\rm R}$ can also be interpreted as auxiliary density matrices. For instance, hermiticity of $\rho_\infty,\rho'_\infty$ is implied by $\ell_{j,j'}=\bar{\ell}_{j',j}$,
$r_{j,j'} = \bar{r}_{j',j}$.
Fixed point conditions (\ref{eq:fixpoint}) trivially follow by telescoping the bulk relation (\ref{eq:bulk}), if the additional boundary equations are satisfied:
\begin{eqnarray}
&&\bra{\rm L}\left(\mathbb L^+_{\mathfrak{ab},1}-
\sum_\mu K_{\rm L,1}^{[\mu]\dagger} \mathbb L^{-}_{\mathfrak{ab},1} K_{\rm L,1}^{[\mu]}\right) = 0\,, \label{eq:LBE}\\
&&
\left(\mathbb L^-_{\mathfrak{ab},N}-
\sum_\mu  K_{{\rm R},N}^{[\mu]\dagger} \mathbb L^+_{\mathfrak{ab},N} K_{{\rm R},N}^{[\mu]}\right)\ket{\rm R} = 0\,.
\label{eq:RBE}
\end{eqnarray}
For example, the left, and right boundary equation (\ref{eq:LBE},\ref{eq:RBE}) should fix $\ell_{j,j'}$, and $r_{j,j'}$, respectively.
We assume that an exact solution should exist within a truncated auxiliary space considering only $J_{\rm L}$ states 
$\{\ket{j}_{\mathfrak{a},\mathfrak{b}};j=0,1\ldots,J_{\rm L}-1\}$ at the left boundary. Due to band triangular structure of Lax operators (\ref{eq:Ls}) one has to consider $J_n=J_{\rm L}+n$ auxiliary states at site $n$ and finally $J_{\rm R}=J_{\rm L}+N$ states at the right boundary. 

\emph{Easy plane regime.---} In order to proceed towards explict solutions of the above equations, we have to specify the magnetic regime of the unitary six-vertex model (\ref{Ugate}). We start with the EPR, $|q|=1$ and $\lambda\in\mathbb R$, implying 
$$A^{[n]\dagger} = \begin{pmatrix} 1 & q^{-n}\bar{z} \cr
-q^n/\bar{z} & -1\end{pmatrix}, \; D^\dagger = D,\; E^\dagger = E.$$
We find explicit solution of the 
left boundary equation (\ref{eq:LBE}) for $J_{\rm L}=2$, exactly when $z$ equals to the polarization of the left target state in (\ref{eq:TS}), unique up to a scale:
\begin{eqnarray}
\ell_{0,0}&=& \frac{\lambda+\lambda^{-1}}{2} + \frac{\lambda-\lambda^{-1}}{2}\frac{|z|^2-1}{|z|^2+1},\;\;
\ell_{0,1}=\ell_{1,0}=1,\nonumber \\
\ell_{1,1} &=& \frac{\lambda+\lambda^{-1}}{2} - \frac{\lambda-\lambda^{-1}}{2}\frac{|z|^2-1}{|z|^2+1}\,.
\label{eq:lEPR}
\end{eqnarray}
The right boundary equation (\ref{eq:RBE}) then amounts to linear recursion relations for $r_{j,j'}$, which we can solve in the closed form, again uniquely up to overall factor:
\begin{equation}
r_{j,j'} = c_{j,j'} \prod_{k=0}^{j-1} b_{k-M} \prod_{k=0}^{j'-1} \bar{b}_{k-M},\;
b_n = \frac{\lambda z q^{-n} \!-\! w q^n}{z q^{-n-1}\!-\!\lambda w q^{n+1}}, \label{eq:RBV}
\end{equation}
where $j,j'= 0,1\ldots N+1$, $M=(N+1)/2$ is an integer, and
\begin{equation}
c_{j,j'} = (-1)^{j-j'} \left(|z| q^{j'-j} + |z|^{-1}q^{j-j'}\right)\,. \label{eq:c}
\end{equation}
Note that both, left and right boundary vectors are generally non-separable w.r.t. partitioning of a pair of auxilliary spaces $\mathfrak{a}\otimes\mathfrak{b}$.
In fact, the Schmidt rank for both boundary vectors is equal to $2$ (for the right boundary this follows from presence of two terms in expression (\ref{eq:c})).
Thus 
(\ref{eq:MPA}) cannot be factorized into a product of two Cholesky-type 
factors as it was the case in all 
the previous solvable instances ~\cite{TopicalReview,ZadnikPRL2020,ZadnikPRE2020,PopkovPRB2022}.
%Note that  $\ell_{j,j'}\neq f_j g_{j'}\ \  \forall j,j'$, and consequently
%the two copies of the auxiliary 
%spaces present in (\ref{eq:MPA}) are coupled. 

\emph{Easy axis regime.---}
In EAR, $q\in\mathbb R$ and $|\lambda|=1$, we have the following conjugated matrices
$$
A^{[n]\dagger} = \begin{pmatrix} 1 & q^{n}\bar{z} \cr
-q^{-n}/\bar{z} & -1\end{pmatrix},\; D^\dagger = E/\lambda,\; E^\dagger = D/\lambda,$$
entering the construction of double Lax operators $\mathbb L^\pm_{\mathfrak{ab},n}$.
Following analogous steps as for EPR we again find explicit non-separable solution for
$J_{\rm L}=2$, namely:
\begin{equation}
\ell_{0,0}=\ell_{1,1}=1\,\quad
\ell_{0,1}=\bar{\ell}_{1,0}=\frac{\lambda|z|+
\lambda^{-1}|z|^{-1}}{|z|+|z|^{-1}}\,,\\
\label{eq:lEAR}
\end{equation}
and the right boundary vector given by exactly the same expression (\ref{eq:RBV}) but $c_{j,j'}$ modified to
\begin{equation}
c_{j,j'}=(-1)^{j+j'}\left(|z| q^{2M-j-j'} + |z|^{-1} q^{j+j'-2M}\right)\,.
\end{equation}

\begin{figure}[tbp]
	\centering
	\includegraphics[width=0.8\columnwidth]{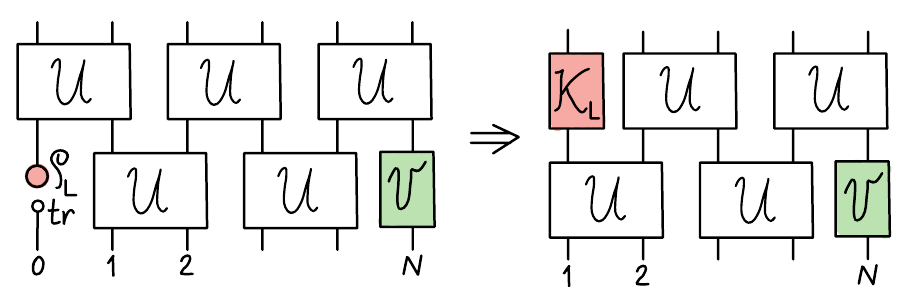}
	\caption{Hybrid driven $XXZ$ circuit in folded notation where left boundary is driven by reset channel and right boundary by local unitary gate. Note that total number of physical qubits $(N+1)$ is now one less than in two-reset setup Fig.~\ref{fig:scheme}.}
	\label{fig:scheme2}
\end{figure}

\emph{Hybrid (coherent-incoherent) boundary driving.---}
Here we show that a similar structure of NESS exists for a hybrid boundary driven setup with the reset channel $\mathcal K_{\rm L}$, Eq.~(\ref{eq:reset}) on the left, while the right-end qubit $n=N$ is driven by an arbitrary local unitary channel (Fig.~\ref{fig:scheme2})
\begin{equation}
\mathcal V(\rho) = V \rho V^\dagger,
\;\,
V = \frac{1}{\sqrt{1+t^2}}
\begin{pmatrix}
(u v)^{-1} & -u^{-1}v t \cr
u v^{-1} t & u v
\end{pmatrix},
\end{equation}
where $u=e^{{\rm i}\alpha}$, 
$v=e^{{\rm i}\gamma}$, $t=\tan \beta$, $\alpha,\beta,\gamma\in\mathbb R$ being the standard Euler angles. 
We postulate that the NESS fixed point of this problem is given by the same ansatz  (\ref{eq:MPA}).  The left boundary equation (\ref{eq:LBE}) holds, so we retrieve the same solution for the left boundary vector (\ref{eq:lEPR}) or (\ref{eq:lEAR}), in the EPR or EAR, respectively. 
The right boundary equation (\ref{eq:RBE}), however, needs to be replaced by
\begin{equation}
\left(\mathbb L^-_{\mathfrak{ab},N}-
V_N \mathbb L^+_{\mathfrak{ab},N} V^\dagger_N\right)\ket{\rm R} = 0\,.
\label{eq:RBE2}
\end{equation}
%\textit{Again, using MPA (\ref{eq:MPA})
%we find again the same solution, 
%(\ref{eq:lEPR}) or (\ref{eq:lEAR}), in the EPR or EAR, %respectively. }
Remarkably, the
solution  of the recurrence scheme for $r_{j,j'}$ following from (\ref{eq:RBE2}) is now identical in both regimes (EPR/EAR) and reads:
\begin{eqnarray}
r_{j,j'} &=& (-1)^{j-j'}\prod_{k=0}^{j-1} g_{k-M} \prod_{k=0}^{j'-1} \bar{g}_{k-M}\,,\label{eq:RBV2}\\
g_n &=& \frac{
(1-v^2 z t q^{-2n-1})z \lambda -
u^2(t q^{2n+1} + v^2 z)}
{(1-v^2 z t q^{-2n-1}) z -
u^2 (t q^{2n+1} + v^2 z)\lambda}\,,
\nonumber
\end{eqnarray}
where $j,j'=0,1\ldots N+1$ and $M=(N+1)/2$.
Note that the product (separable) form of the boundary vector (\ref{eq:RBV2}) is a consequence of unitarity of the right boundary channel. This solution partly generalizes a related setup in the continuous-time Lindbladian case and transverse field~\cite{Clerk2024}.

%%%%%%%%%%%%%%%%%%%%%%

\emph{Spin helices in brickwork circuits.---}~
Simple analysis  of zeros and poles of the coefficients $b_n$ in (\ref{eq:RBV}) leads to prediction of remarkably simple pure and separable NESS:
A specific  choice of the right boundary reset 
\begin{equation}
w=q^{N+1} z \lambda\,,
\label{SHScond}
\end{equation}
corresponding to $b_{-M}=0$, leads to 
$r_{i,j}= \delta_{i,0}\, \delta_{j,0}$ and to separable NESS, 
 $\rho_\infty =\ket{\Psi^+_{\!S}}\!\!\bra{\Psi^+_{\!S}}$,
 $\rho'_\infty =\ket{\Psi^-_{\!S}}\!\!\bra{\Psi^-_{\!S}}$:
\begin{eqnarray}
&\ket{\Psi^\pm_S }= \ket{\psi_1^\pm} \otimes \ket{\psi_2^\mp} \otimes \ket{\psi_3^\pm}\otimes\ket{\psi_4^\mp}\cdots\ket{ \psi_N^\pm}, \label{SHS}\\
&\ket{\psi^{+}_n} =
\begin{pmatrix}
    1\cr z q^n
\end{pmatrix}, \quad \ket{\psi^{-}_n} = 
\begin{pmatrix}
    1\cr \lambda z q^n
\end{pmatrix}.  \label{SHS1}
\end{eqnarray}
%\begin{align}
%&\ket{\Psi_S }= \psi_1' \otimes \psi_2  \otimes \psi_3'  \otimes %\psi_4 \ldots   \otimes \psi_N'   \label{SHS}\\
%&\psi_n = \binom{1}{z q^n}, \quad \psi_m' = \binom{1}{ z \lambda %q^m},  \label{SHS1}
%\end{align}
%
%which amounts to 
%\begin{align}
%&\rho_\infty = \prod_{n=1}^{(N+1)/2} E_{2n-1} A_{2n-1}^{[2n-1]}
%A_{2n-1}^{[2n-1]}^\dagger E_{2n-1} \nonumber \\
%&\times \prod_{n=1}^{(N-1)/2}  A_{2n}^{[2n]}D_{2n}^2
%A_{2n}^{[2n]}^\dagger, 
%\end{align}
%which is a pure state,  $\rho_\infty =\ket{\Psi_S }\bra{\Psi_S }$, 
%where 
%
%
%\begin{align}
%&\ket{\Psi_S }= \psi_1' \otimes \psi_2  \otimes \psi_3'  \otimes %\psi_4 \ldots   \otimes \psi_N'   \label{SHS}\\
%&\psi_n = \binom{1}{z q^n}, \quad \psi_m' = \binom{1}{ z \lambda %q^m},  \label{SHS1}
%\end{align}
In EPR the arrangement of qubits in the state (\ref{SHS}) represents a distorted helix with 
wavelength $\Lambda=(2\pi i)/(\log q)$: the azimuthal angle of the helix $\varphi(n) = \eta n $, $ \eta = -i \log q$
grows linearly with the position $n$,  while the polar angle exhibits even/odd staggering with values $\theta_L, \theta_R$,
where $\tan\frac{\theta_R}{2} = \lambda \tan\frac{\theta_L}{2}$. 
In the continuous time limit, $\lambda \rightarrow 1$, the state (\ref{SHS}) becomes a perfect spin helix.  
With some abuse of notations we shall refer to  (\ref{SHS}) as a  \textit{brickwork helix} or just \textit{helix} state.
Like their continuous time counterparts brickwork helices are very atypical chiral state with non-zero size-independent spin-current.

Analogously,  a pole in $b_{N-M}$ 
($1/b_{N-M}=0$) yields a helix with inverted helicity,  namely,  a choice
\begin{align}
&z= w\lambda q^{N+1} \label{SHSinverted}
\end{align}
yields again a pure NESS, of exactly the same form as (\ref{SHS})
with redefinitions $\ket{\psi^{+}_n}= \binom{1}{zq^{-n}}$, $\ket{\psi^{-}_n}= \binom{\lambda}{zq^{-n}}$.

Significance of the distorted spin helix  (\ref{SHS}) and its counterpart with opposite helicity
for the circuit (Fig.~\ref{fig:scheme}) lies beyond just 
being  exact NESS for the special choice of the right boundary conditions (\ref{SHScond}) or (\ref{SHSinverted}).
It lies in its robustness: indeed 
 due to the Yang-Baxter relation (\ref{eq:bulk}) the helix is reproduced 
locally in the bulk after every cycle.  Correspondingly, 
regardless of 
conditions  imposed on the right boundary the helix after one cycle will only be changed at the two rightmost sites
while everywhere else it will stay intact.  It requires a time of order of system size $N$ to destroy the
helix completely while typical state will be destroyed after one time step. Being thus  exceptionally robust, and in addition, easily creatable (being a
factorized state),  the helix can be used for calibration purposes,  similarly to its 
facility in the continuous time counterpart $\lambda \rightarrow 0$ in cold atoms \cite{KettNature2022}.

%Note also that  the state (\ref{SHS}) carries exceptionally %large spin current, which is system-size independent.

A slight modification of condition (\ref{SHScond}), 
\begin{align}
w=q^{N+1-2 m_{\rm k}} z \lambda,
\label{KinkCondition}
\end{align}
with integer $m_{\rm k}=1,2,\ldots$ leads to cropping of the right boundary vector to dimension $(m_{\rm k}+1)^2$,
$r_{j,j'}=0$ if $j>m_{\rm k}$ or $j'>m_{\rm k}$.
The respective NESS is a
mixed state of rank $(m_{\rm k}+1)^2$, and can be viewed as consisting from  imperfect helices with $m_{\rm k}$ kinks (in analogy to~\cite{PopkovSHS2}). 
For helices with inverted helicity the analogue of  (\ref{KinkCondition}) reads 
$z/w=\lambda q^{N+1-2 m_{\rm k}} $, generalizing (\ref{SHSinverted}).

To detect the helix states,  
instead of measuring the NESS magnetization current, which is a complicated 4-point correlation 
for a brickwork XXZ circuit, see \cite{LjubotinaPRL19}, we propose to measure instead a 
simple one-point 
correlation,  
$\langle \sigma_1^+ \rangle = 
\tr(\sigma^+_1\rho_\infty )$,  
which turns out to be extremely sensitive
to the presence of  helices, and their descendants (helices with kinks). 
From complex $\langle \sigma_1^+ \rangle $, we construct two scalar helix indicators 
\begin{align}
&f_{1} = 1-  \frac {\arg(\langle \sigma_1^+ \rangle/z)}{\eta},  \label{def:f1}\\
&f_{2} = -1+\left|  (|z|+ |z|^{-1}) \langle \sigma_1^+ \rangle \right|, \label{def:f2}
\end{align}
$f_1$ relating to the twisting angle between the first qubit and the left reset state,
and $f_2$ relating to purity of the first qubit.  Both  indicators vanish
$f_1=f_2=0$ for the perfect brickwork helix (\ref{SHS}).

 \begin{figure}[tbp]
  \centering
  \includegraphics[width=0.5\textwidth]{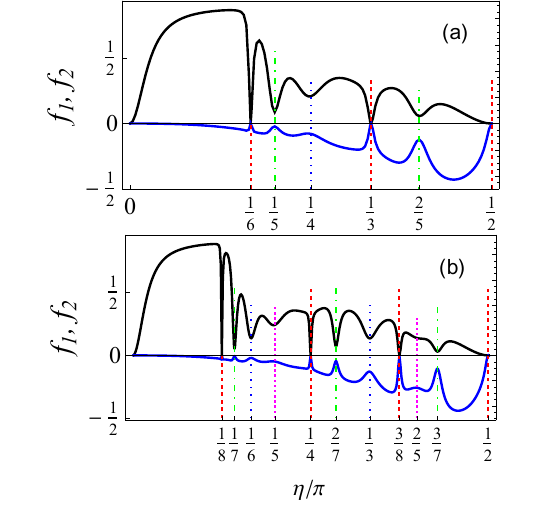}
  \caption{Helix indicators: $f_1$  (\ref{def:f1}) (upper black curve), $f_2$
(\ref{def:f2})  (lower blue curve) measured in the NESS, in the 
easy plane regime, 
versus the XXZ anisotropy $\eta/\pi$, for system of $N=11$ (a) and 
$N=15$ (b)  sites. 
Parameters: $z=1,\lambda =\exp[0.9], w= z \lambda  $.  
Zeros of the observable coincide with the pure helix condition 
$(N+1)\eta = 0\pmod{2\pi}$, 
(red dashed lines), while other less pronounced minima  occur 
at the anisotropy values  leading to  helix with integer number of kinks
$(N+1-2 m_k)\eta = 0\pmod{2\pi}$, $m_k=1,2,3$ (green, blue 
and magenta dashed lines, respectively).
   }
  \label{FigHelixN11}
\end{figure}

According to (\ref{SHScond}), (\ref{SHSinverted}) we expect the helix (\ref{SHS}) to  appear at fine-tuned values of anisotropy 
$\eta = -i \log q$ provided that the boundary reset states satisfy
\begin{align}
&|w/z|=\lambda \quad \mbox{or}\ \ |w/z|=1/\lambda.  \label{ResonanceCondition}
\end{align}
%Eq. (\ref{ResonanceCondition})
%can be viewed as a necessary condition for the helices existence.

 In Fig.  ~\ref{FigHelixN11} we plot  $f_1$ and $f_2$ versus the anisotropy $\eta$ in a chain 
with reset channels fulfilling  (\ref{ResonanceCondition}).
Apart from the expected zeros of $f_1,f_2$ signalizing pure NESS,  we  find sharp local minima of $f_1$ and $-f_2$, 
exactly at the $\eta$ values (\ref{KinkCondition}) where we predicted helix descendants to occur.
The  number of pure helices and their descendants increase linearly with the system size
$N$ and, in addition, for larger $N$ more and more descendants are resolved in $f_1(\eta), f_2(\eta)$ dependence. 
This can be seen comparing the upper and lower panels of Fig. ~\ref{FigHelixN11}.
Indeed for $N=11$ one easily identifies minima corresponding to $m_{\rm k}=0,1,2$ in (\ref{KinkCondition}) while for $N=15$ one can  also
recognize the minima corresponding to states with three kinks $m_{\rm k}=3$.
In absence of the resonant condition 
(\ref{ResonanceCondition}) the resonance peaks are absent and the local chirality of the NESS ($1-f_1$) decreases with
systems size $N$ irrespectively of the anisotropy (data not shown).

Hybrid systems (Fig.~\ref{fig:scheme2}) exhibit similar phenomenology and even larger sensitivity to the presence of
helices. 

\emph{Discussion.---} 
We have presented an exact solution for the nonequilibrium steady state of brickwork circuits in two cases:
(i) with two reset channels at the boundaries, and (ii) one reset channel at one boundary and local field at another boundary.
The system we considered can be directly realized experimentally,  with technique belonging to the 
experimental toolbox \cite{GQAI_Science24a,GQAI_Nature22}.
We also identified the simplest yet nontrivial steady states of the driven brickwork circuits, brickwork helices
(\ref{SHS}), which should be of
direct experimental relevance being especially robust and easily manipulatable: these states feature qubits,
arranged  as helices with even-odd site staggering.  The  period of the helix,  $\Lambda=2  \pi i/\log[q]$,
 fixed by the anisotropy parameter  $q$  of the principal bulk gate (\ref{Ugate}), is system size independent. 
 Brickwork helices and their descendants,  helices with kinks (\ref{KinkCondition}) are identifiable
 by measuring just a  one-point correlation in the steady state as exemplified in  Fig.~\ref{FigHelixN11},  which  again can be of direct relevance for an experiment.

 The technical framework introduced here should be directly applicable to other reset-driven brickwork circuits with local gates satisfying the (braid) Yang-Baxter equation.
 
%\begin{acknowledgements}
  V.P. and T.P. acknowledge support by ERC Advanced grant
  No.~101096208 -- QUEST, and Research Programme P1-0402 of Slovenian
  Research and Innovation Agency (ARIS). V.P. is also supported by
  Deutsche Forschungsgemeinschaft through DFG project
  KL645/20-2. 
%\end{acknowledgements}
  
\bibliography{bibfile.bib}
 
\end{document}